\documentclass[graybox]{svmult}

\usepackage{mathptmx}       
\usepackage{helvet}         
\usepackage{courier}        
\usepackage{type1cm}        
\usepackage{makeidx}         
\usepackage{graphicx}        
\usepackage{multicol}        
\usepackage[bottom]{footmisc}

\makeindex             

\def\Msol{$\rm M_{\odot}$}
\def\mwlike{\mbox{$\mathcal{M}_{h12}$}}

\begin{document}

\title*{The Wide Area VISTA Extra-galactic Survey (WAVES)}
\author{Simon P. Driver, Luke J. Davies, Martin Meyer, Chris Power, Aaron S.G. Robotham, Ivan K. Baldry, Jochen Liske and Peder Norberg}
\institute{Simon P. Driver, Luke J. Davies, Martin Meyer, Chris Power, Aaron S.G. Robotham \at 
International Centre for Radio Astronomy Research (ICRAR), School of Physics, University of Western Australia, M468, 35 Stirling Highway, Crawley, Western Australia, WA 6009 \email{simon.driver@uwa.edu.au}
\and Ivan K.Baldry \at Astrophysics Research Institute, Liverpool John Moores University, IC2, Liverpool Science Park, 146 Brownlow Hill Liverpool, L3 5RF, UK
\and Jochen Liske \at European Southern Observatory, Karl-Schwarzschild-Str. 2, 85748, Garching, Germany
\and Peder Norberg \at International Cosmology Centre, Durham University, Durham, DH1 3LE, UK
}
%
%
\maketitle

\abstract{The ``Wide Area VISTA Extra-galactic Survey'' (WAVES) is a
  4MOST Consortium Design Reference Survey which will use the
  VISTA/4MOST facility to spectroscopically survey $\sim2$\,million
  galaxies to $r_{\rm AB} < 22$ mag.
  WAVES consists of two interlocking galaxy surveys (``WAVES-Deep''
  and ``WAVES-Wide''), providing the next two steps beyond the highly
  successful 1M galaxy Sloan Digital Sky Survey and the 250k Galaxy
  And Mass Assembly survey. WAVES will enable an unprecedented study
  of the distribution and evolution of mass, energy, and structures
  extending from 1-kpc dwarf galaxies in the local void to the
  morphologies of 200-Mpc filaments at $z\sim1$. A key aim of both
  surveys will be to compare comprehensive empirical observations of
  the spatial properties of galaxies, groups, and filaments, against
  state-of-the-art numerical simulations to distinguish between
  various Dark Matter models.  }

\section{Introduction}
Since the pioneering days of the 2dFGRS and SDSS, extra-galactic
spectroscopic surveys have come in two flavours: those optimised for
cosmology, and those optimised for galaxy evolution. The distinction
is important. Cosmology surveys (e.g., WiggleZ, BOSS, DESI) use
specific tracers to probe the underlying large scale structure. These
surveys advance cosmology but provide a biased cross-section of the
galaxy population. Conversely galaxy-evolution studies (e.g., MGC,
GAMA, zCOSMOS, DEEP2, MOONS) uniformally sample the full galaxy
population, but only cover modest areas. These samples are ideal for
studying galaxy evolution and its interplay with environment, but lack
the area coverage for cosmological studies. WAVES, in the era of
dedicated cosmology experiments, represents the next step in galaxy
evolution studies, bridging the gap between the very near ($z<0.3$;
SDSS, GAMA) and the very far ($z>0.8$; MOONS, HST, JWST), as well as
probing the intrinsically faint (low mass) and dim (low surface
density) populations within the nearby Universe ($z<0.2$).

The WAVES target catalogue will be based on the VST KiDS South
sub-arcsecond optical imaging and is intended to complement Euclid,
LSST, and the SKA. The key science drivers and initial survey concept
design are outlined in the sections which follow. Here we briefly
introduce the two WAVES components (Deep and Wide),
outline the primary science motivations, and describe the
preliminary design concept.

\begin{figure}
\includegraphics[width=\textwidth]{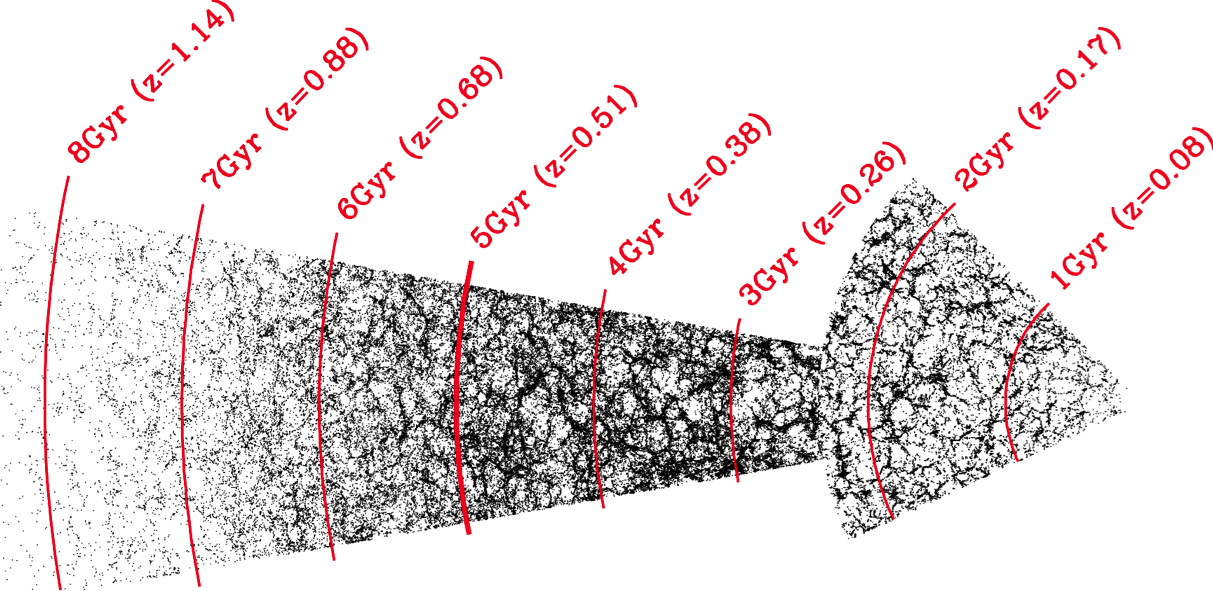}
\caption{A representation of the RA geometry of the WAVES survey
  (derived from the Theoretical Astrophysical Observatory),
  highlighting the complexity of structures that will be sampled.
\label{fig:dwwaves}}
\end{figure}


\noindent  
  {\bf WAVES-Deep} will cover 100 sq.~deg.\ to $r_{\rm AB}\sim 22$~mag
  and extend the power of SDSS and GAMA like population statistics out
  to $z\sim 0.7$, tracing the rarest structures even up to and
  slightly beyond $z\sim1$. 
  This $\sim1.2$\, million galaxy redshifts survey is key in providing
  the largest and most representative galaxy 
  group and filament catalogue ever constructed.


~

\noindent
  {\bf WAVES-Wide} will cover 750 sq.~deg.\ to 
  $r_{\rm AB} \sim 22$~mag with additional photo-z pre-selection
  ($z_{\rm photo} < 0.2$). This will target $\sim0.9$\, million galaxies 
  allowing a detailed study of the occupancy of
  $\sim 10^{11}$--$10^{14}$M$_{\odot}$ halos to a stellar mass limit of
  $10^{7}$M$_{\odot}$, and providing a dwarf galaxy sample over a
  representative volume of $10^6{\rm\,Mpc}^3$.  

\section{WAVES Science Drivers}

\subsection{Ensemble of Milky-Way sized systems to test CDM}

\begin{figure}
\includegraphics[width=\textwidth]{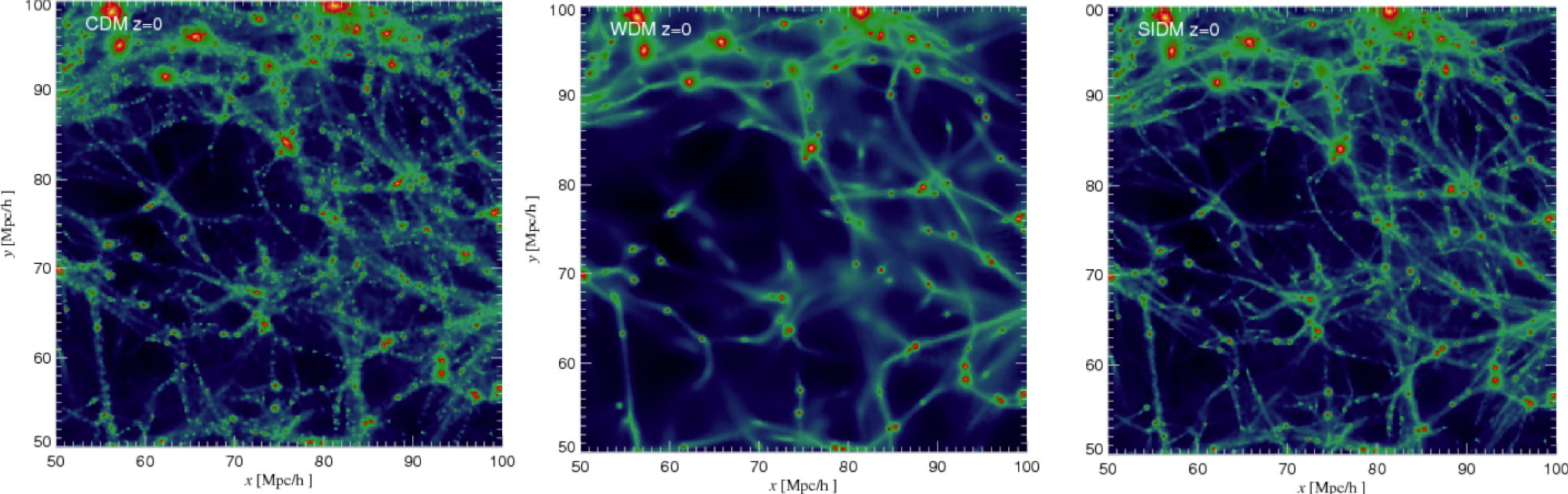}
\caption{ \baselineskip=0pt \footnotesize Numerical simulations of the
  galaxy distribution at z=0 from (left-to-right) Cold, Warm, and
  Self-interacting dark matter. The images show the dark matter
  density within a cube of
  $50$Mpc$\times 50$Mpc$\times 5$Mpc \label{fig:cdmwdmsidm}}
\end{figure}

The nature of dark matter is one of the key questions in modern day
cosmology. The currently favoured Cold Dark Matter model,
$\Lambda$CDM, provides a good description of the large scale structure
of the Universe (see Figs.~\ref{fig:dwwaves}
\&~\ref{fig:cdmwdmsidm}). Comparison of robust model predictions with
empirical galaxy clustering measurements on Mpc scales supports the
Cold Dark Matter model for the growth of structure
(e.g. \cite{zehavi11}). On sub-Mpc scale, i.e. on galaxy and group
scales, baryons and baryonic physics become critical: the kpc to Mpc
range is the key scale over which Dark Matter halos virialize and
merge, and baryons decouple, collapse and eventually form complex
structures such as galaxies. In this regime, our theoretical
understanding is less well-founded, in great part due to the immense
complexity of the physics encountered.

%
Galaxy group samples are now able to probe down to a few
$10^{12}$\Msol, with arguably the most complete being the ``GAMA
Galaxy Group Catalogue'' (G$^3$C; \cite{robotham11}).
Properties of low mass galaxy groups are limited by the intrinsic
lack of survey depth.
%
This explains why for $10^{12}$\Msol\ systems (\mwlike\ hereafter),
only two really well studied examples exist: the Milky-Way and the
Andromeda systems, both in our own neighbourhood. 
Galaxies in the Local Group are extensively used to probe in detail
the Cold Dark Matter paradigm. They are often central to the strongest
evidence against the standard Cold Dark Matter model, from ``the
missing satellite problem'' (e.g.\cite{moore99}) to ``the too big
too fail problem'' (e.g.\cite{boylan-kolchin11}).  via ``the
co-planar location of satellite galaxies (e.g.\cite{ibata13}).
Given that \mwlike\ halos are the most important ones in
terms of galaxy formation (with galaxy formation efficiency peaking
just  around that mass scale in all standard CDM model)
and one of the more critical in terms of testing CDM, it is
fundamental to ensure that this limited sample of well studied 
\mwlike\ systems is representative. To create such statistical
sample is a central goal of WAVES. 

WAVES-Wide, which survey depth results in LMC like galaxies to be close to
volume limited out to $z\sim0.2$, is specifically designed to deliver
a high fidelity group catalogue ($\sim$4k groups with 5 or more
members) probing to the very lowest halo masses. This sample size
should allow the intrinsic scatter in the sub-halo / stellar mass
occupation statistics to be measured. WAVES-wide will result in a
proper characterisation of \mwlike\ groups, including assessing how
representative our two best studied examples are. 

\subsection{The low surface brightness and dwarf domains}
The study of the field dwarf galaxy population offers a unique testing
ground for galaxy formation and the underlying physics. Cold dark
matter (CDM) simulations predict that there are many more low-mass
than high-mass halos remaining today (\cite{moore99}).
Observationally there appears to be a deficit with respect to this
prediction; the observed galaxy stellar mass function is not as steep
as the halo mass function. This could be in part because the mass of
the dark matter particle is in the keV range (i.e., WDM), which
suppresses power on dwarf galaxy scales (\cite{bode01}).  However,
dwarf galaxy formation is also sensitive to the impact of the
photo-ionizing background, supernovae feedback, and environmental
effects (\cite{benson03}). Disentangling the baryonic effects from
any change in the power spectrum requires large statistically
representative samples.

\begin{figure}
\includegraphics[width=\textwidth]{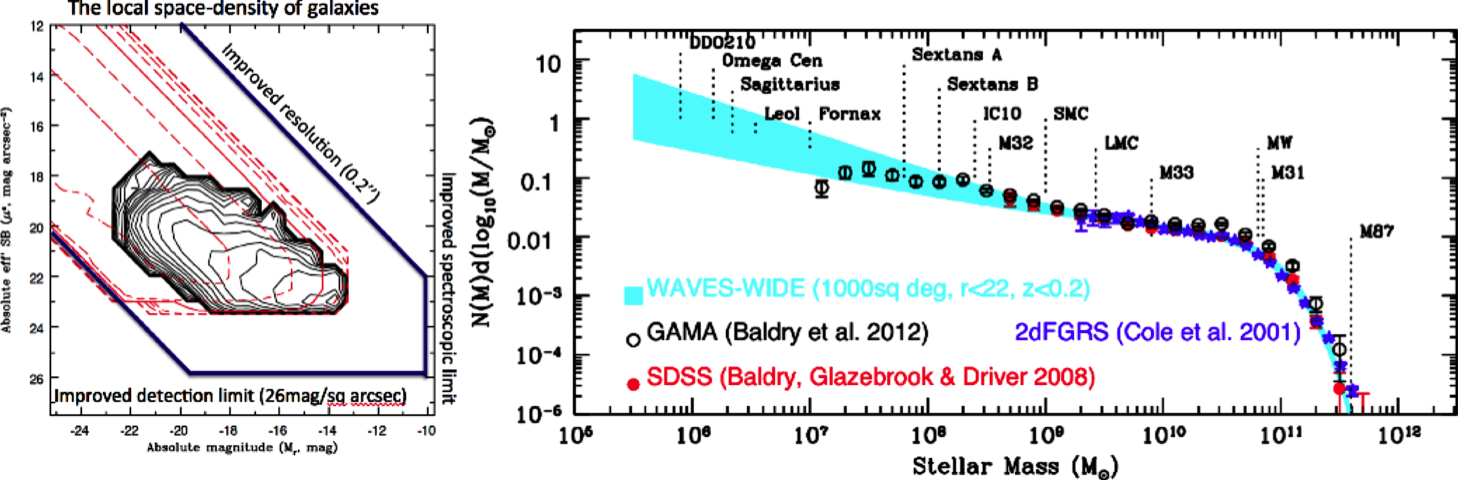}
\caption{\baselineskip=0pt \footnotesize (left) The space-density of
  galaxies in the luminosity-surface brightness plane (derived from
  GAMA). The logarithmic density contours in red show the current
  robust sampling region bounded by the selection limits of the Sloan
  Digital Sky Survey. WAVES will extend our census of the nearby
  galaxy population to the blue boundary due to the superb imaging
  quality and limiting surface brightness sensitivity of the VST KiDS
  data. (right) The improvement in the measurement of the stellar mass
  function possible with WAVES-Wide. \label{fig:smass}}
\end{figure}

Local surveys such as the SDSS and follow-on campaigns such as GAMA
are fundamentally limited by surface brightness sensitivity of the
imaging data, and the reality of the luminosity-surface brightness
relation (\cite{choloniewski85},\cite{dejong00}). Various studies
including \cite{blanton05,driver05,baldry12} clearly demonstrate that
the SDSS becomes incomplete for systems with $\mu_e \geq 23.0$ mags/sq
arcsec which become frequent below absolute magnitudes of $M_r=-18$
mag (Fig.~\ref{fig:smass} left). For this reason the low surface
brightness and dwarf galaxy domains continue to remain uncharted
territory for modern surveys. Only deep surveys of clusters (highly
unrepresentative of the average Universe) have entered this dwarf
galaxy low-SB regime. The first deep wide-area imaging survey capable
of probing into this domain will be VST KiDS covering 750 sq. deg in
the Southern and Northern Galactic Caps. WAVES will use the VST KiDS
data as its input survey to provide targets to $r_{\rm AB} < 22$mag
with $\mu_{e} < 26$ mag/sq arcsec allowing for the construction of a
complete sample of galaxies to $M_r=-14$mag, i.e., 4mags deeper than
SDSS (see the blue boundary in Fig.~\ref{fig:smass} left).  The WAVES
survey, $r_{\rm AB} < 22$, with high-completeness for low-SB galaxies
would enable the galaxy stellar mass function to be measured
accurately down to $10^6$M$_{\odot}$ (see Fig.~\ref{fig:smass} right).

\subsection{The evolution of galaxy structure (with Euclid)}
\begin{figure}
\includegraphics[width=\textwidth]{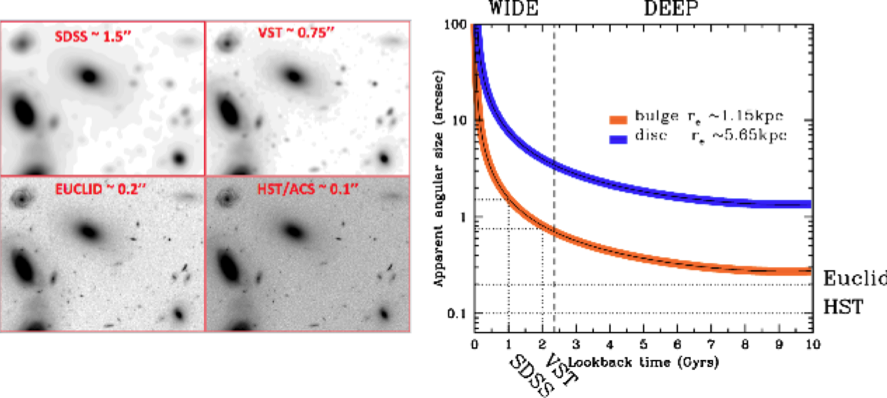}
\caption{ \baselineskip=0pt \footnotesize Angular-size versus lookback
  time with the low-z limitations of the SDSS and VST shown. (right)
  Comparison of various imaging datasets, with median seeing as
  indicated. \label{fig:images} }
\end{figure}

Studies of the mass-size relation of galaxies in the nearby ($z<0.1$)
and distant Universe ($z>1$) show a $\times 5$ growth in galaxy sizes
at fixed stellar mass (e.g., \cite{trujillo06,vandokkum08}) and in
number-density (e.g., \cite{faber07}). This extraordinary result,
confirmed by numerous groups (e.g., \cite{bruce12}), implies a
dramatic physical change occurring in the galaxy population over the
redshift range $0.2-1.0$. Possible explanations include dynamical
relaxation, major mergers, minor mergers, and disc growth
(\cite{driver13}). However this result is only clearly established for
high stellar mass systems (i.e., $>10^{11}M_{\odot}$), found in
extreme dense cluster environments. Whether this growth is endemic or
confined to a specific mass or environment remains unclear. A deep
spectroscopic survey with HST resolution imaging over a sustained area
is needed to extend the measurements to fainter mass limits and to
distinguish between the competing hypotheses. WAVES-Deep in
combination with Euclid imaging (see Fig.~\ref{fig:images} left),
provides exactly the dataset required to study this extraordinary
growth.

A key starting point will be to address whether the fundamental nature
of galaxies is its bimodality (red v blue), or its evident duality
(bulges plus discs).  At very low redshift ($z<0.1$) this issue is
clear-cut with multi-component decomposition a routine part of the
analysis toolkit. At high redshift ($z>1.5$) the case is less clear as
galaxies no-longer appear to adhere to the simple idea of bulge plus
disc systems, but exhibit highly asymmetrical and irregular shapes. As
such the language of high-redshift galaxy work is typically focused
on turbulence, distortions, and the global colour (red v blue). 

To date the largest contiguous survey by HST covers 1.8 sq deg
(COSMOS). Euclid will transform this by imaging upto 8000sq degrees of
sky at 0.2$''$ resolution.
Surveying this entire area spectroscopically is unrealistic, however
WAVES-Deep will provide over 1-million galaxies in the range $0.2 < z
< 0.8$ with imaging resolution sufficient to discern and measure
bulge, bar, and disc components to 1kpc resolution (see
Fig.~\ref{fig:images} right). This will open the door for direct
measurements of the mass and size evolution of the distinct structural
components (bulges, bars, discs) within a sufficiently comprehensive
survey to dissect trends by halo mass, star-formation rate or a
multitude of other indicators. {\it If} tracing the duality of
galaxies is critical for understanding galaxy formation, as we suspect
(\cite{driver13}), then WAVES-Deep/Euclid will provide more than an
order of magnitude advancement over COSMOS/zCOSMOS, and in doing so
firmly bridge the near and far Universe.

\subsection{The evolving HI universe (with ASKAP/SKA)}

The SKA and its pathfinders will allow, for the first time, a direct
study of the HI content of galaxies and its role in galaxy evolution
over cosmologically representative volumes and over significant cosmic
time.  Understanding the complicated interaction between the gas
content of galaxies, their environments, and other major galactic
constituents, remains a largely unsolved problem.  In anticipation of
the next generation of observational programs, current simulations of
cosmic evolution now increasingly provide measures of gas content and
its phase breakdown (atomic, molecular, ionized; e.g.,
\cite{obreschkow09, power10, lagos11}), and numerous models have been
advanced for the fundamental physical processes that underly the
observable gas scaling relations.  These models must now be tested,
and our observational understanding of evolution in the scaling
relations improved to advance the field.

The advantages of alignment between HI and optical spectroscopic
programs are many-fold: a comparison of gas content with high fidelity
measures of environment and halo mass from optical spectroscopy,
metrics that are imperfectly traced by HI galaxy redshifts alone due
to the strongly anti-biased nature of this population; improved
treatment of issues such as galaxy confusion, satellites, and
multi-wavelength counterpart identification due to the comparatively
higher spatial resolution of optical data; optically motivated source
finding to increase the size of HI source catalogues; and the maximal
exploitation of deep HI data through the application of statistical
techniques, such as HI stacking, that increase the cosmic baseline
over which evolutionary measurements of gas content can be made.
Optical redshift catalogues further enable the properties of the
gas-rich population to be directly compared to the general galaxy
population as selected by tracers much more closely linked to stellar
and halo mass.

While the current generation of optical redshift surveys are
well-suited to the SKA pathfinders, such datasets are absent for SKA
phase 1.  In particular, SKA1-mid is expected to survey a ~100sq deg
region (i.e., comparable to WAVES-Deep) out to $z \sim 1$ and beyond.
No such dataset currently exists, being too expensive in telescope
time for current facilities, and a unique opportunity for VISTA that
will be met by WAVES.  Combined with data from SKA phase 1 and Euclid,
Deep-WAVES will enable robust measurement of HI over half the history
of the Universe, and unique measurements over cosmologically
representative volumes of the complex interplay of gas content with
other major galactic constituents.

\subsection{A legacy resource}
\begin{figure}
\includegraphics[width=\textwidth]{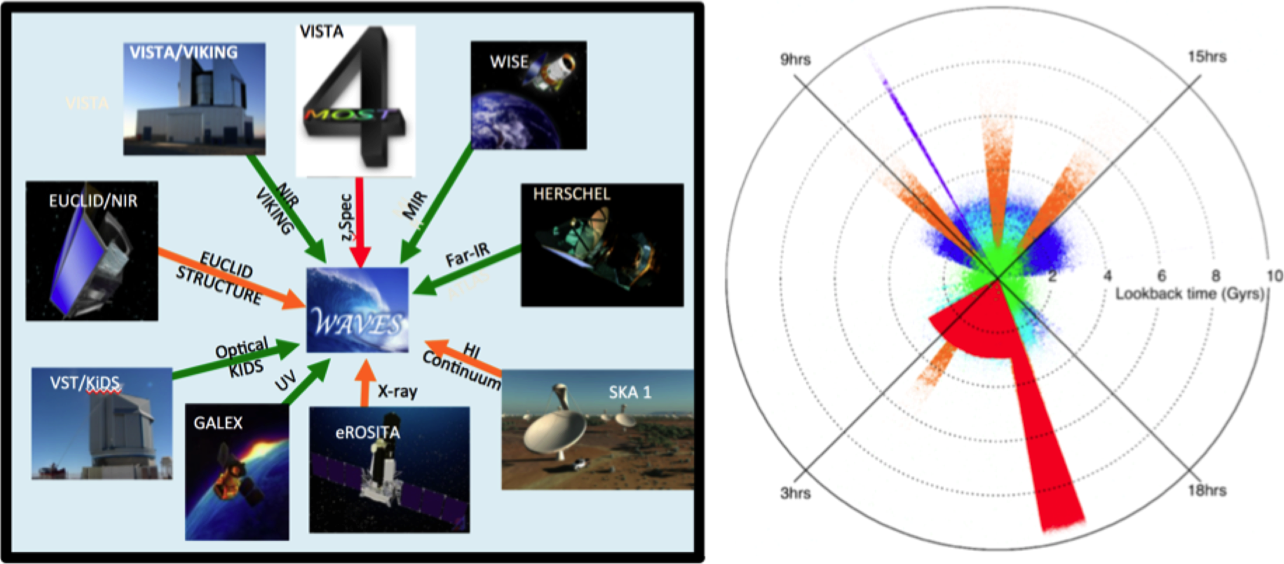}
\caption{(left) The facilities ultimately contributing to the WAVES
  survey. \label{fig:tels} (right) a cylinder-plot showing the region
  of the Universe probes by WAVES compared to other notable surveys.}
\end{figure}

WAVES is designed as a legacy survey from which numerous science
questions can be asked and numerous follow-on campaigns launched. For
this reason the survey selection, crucial for establishing legacy
value, is kept as simple as possible with only flux selection for
WAVES-Deep and flux and photo-z selection for WAVES-Wide. More complex
selection ultimately minimises the generic usefulness of a survey. For
example the very specific selection algorithms of large cosmology
programmes (e.g., BOSS) reduces the usability of such surveys for
evolutionary studies of the galaxy population. There is also a direct
opportunity to connect the WAVES survey to planned campaigns
(Fig.~\ref{fig:tels}) with Euclid --- capable of providing 0.2$''$
images over the entire WAVES region in optical and near-IR bands ---
SKA Phase I --- capable of sampling HI to $z \sim 0.2$ over WAVES-Wide
and to $z \sim 1.0$ over the WAVES-Deep region as well as the
all-hemisphere/sky surveys to be conducted by LSST and eROSITA. Apart
from fulfilling the science cases outlined above this presents a
unique opportunity to combine a deep spectroscopic campaign with both
Euclid, SKA Phase I, LSST and eROSITA to study the late-time assembly
and evolution of galaxies over a broad mass and redshift baseline with
robust stellar masses, gas masses, morphologies and structural
decompositions.

Galaxy evolution is complex, and the current picture murky. Clarity
will come from comprehensive studies based on the highest quality
inputs combined with rigorous analysis such as that provided by the
Sloan Digital Sky Survey (\cite{york00}) and Galaxy And Mass
Assembly (\cite{driver11}) teams.  WAVES will game-change by
extending to flux limits 4 magnitudes fainter than SDSS locally, and
outward in redshift ($z \sim 1.0$). The scale of each of the proposed
surveys are SDSS-like in scale and the 1-million galaxy ball-park has
proven crucial in establishing key trends such as the mass-metallicity
relation (\cite{tremonti04}) and other key measurements. In addition
conducting a fully sampled contiguous survey provides the opportunity
to construct a robust halo mass catalogue to explore the role and
influence of the halo on galaxy evolution (\cite{robotham13}) as
well as the intrinsic properties of halos themselves (\cite{robotham08}).

Hence with WAVES-Deep one can ask questions such as whether the
stellar mass-size relation of spheroids and discs evolves in the same
manner irrespective of halo mass, or how does the stellar-to-gas mass
fraction vary over 4dex in stellar mass from rich clusters, to
filaments, to voids. Perhaps most important of all but not discussed
in detail here are the possibilities opened up by the $R \sim 5000$
mid-resolution spectrograph which will provide robust metalicity, age,
dust and star-formation measurements (e.g., GANDALF; \cite{sarzi06})
for the higher signal-to-noise subset (i.e., $r_{\rm AB} < 21$ mag and S/N/\AA$\sim 20$),
of the WAVES galaxies.  Finally we note that the proposed regions are
or will be pre-surveyed by notable facilities including
GALEX, VST (KiDs), VISTA (VIKING), WISE, and Herschel (Herschel-Atlas).

\begin{figure}
\centerline{\includegraphics[width=10.0cm]{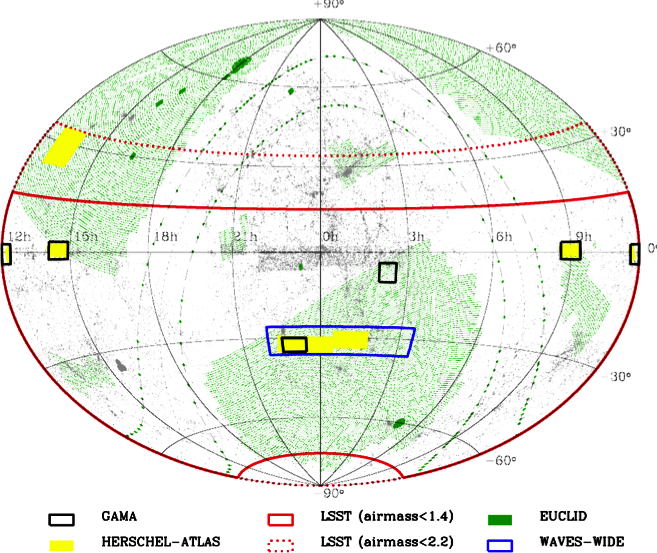}}
\caption{ \baselineskip=0pt \footnotesize An Aitoff projection showing
  the location on the sky of various surveys including GAMA, Herschel-Atlas, Euclid and WAVES-Wide.
\label{fig:aitoff}}
\end{figure}

\section{WAVES survey design}
The survey scope is bounded by the likely capabilities of the
VISTA/4MOST facility, the availability of suitable data to define a
target catalogue, and consideration of available complementary
datasets. Within these bounds the design is then driven by the science
drivers which push towards a comprehensive (100---750 sq deg),
contiguous, faint flux limited ($r_{\rm AB} < 22$ mag) survey. The
tension between studying galaxy evolution over a broad timeline versus
probing down to the faintest halo mass and halo occupation limits
leads to a survey split into WAVES-Deep and WAVES-Wide. For efficiency
and legacy purposes these surveys are interleaved with identical
initial target selection but with an additional photo-z selection
applied to WAVES-Wide (derived from $ugriZYJHK$ VST/VISTA matched
aperture photometry where $\Delta z/(1+z) \sim 0.03$ is
realistic). 

The final WAVES catalogue is expected to contain $\sim 2$\,million
galaxies, 140k groups, and 14k filaments from within the VST KiDS
South footprint (see Fig.~\ref{fig:aitoff}).  This region of sky passes
directly overhead at Paranal and would represent the first major
spectroscopic campaign South of the Galactic equator since 2dFGRS over
a decade ago.

Both WAVES surveys are designed to operate to the limiting sensitivity
of the 4MOST facility, taken here as $r_{\rm AB} < 22$ mag for a 2hr
($5\times 24$min) integrations with the Mid-Res spectrograph ($R \sim
5000$, $\Delta \lambda 0.4-0.9\mu m$, S/N/\AA$\sim3$). All 1800
low-res fibres would be engaged, and the survey density (1000 and 9800
gals/sq.deg for WAVES-Wide and WAVES-Deep) is such that the WAVES-Wide
footprint will be sampled only once while the WAVES-Deep footprint
will be sampled $\sim 6\times$ (ideal for sampling the cores of dense
groups). Pre-selection will consist solely of flux and colours
consistent with galaxies for WAVES-Deep while WAVES-Wide will include
an additional photo-z selected ($z<0.2$) using $ugriZYJHK$ photometry
provided by VST KiDs and VISTA VIKING.

\section{Summary}
As of November 2014 WAVES has been adopted as one of the eight key
Design Reference Surveys for VISTA/4MOST and should commence circa
2021. The WAVES team remains committed to four overriding
principles: (i) simplicity of survey design, (ii) high contiguous
coverage, (iii) maximising the synergy with complementary facilities,
and (iv) the professional release of high quality data products to the
community. Updates on the progress of the project will be broadcast
via the website: http://www.wave-survey.org/

We would like to finish by thanking the organisers for the invitation to
participate in ``The Universe of Digital Sky Surveys'' meeting, our
thanks in particular to Massimo Capaccioli for his vision and resolve
in delivering VST (upon which WAVES builds), and the editors for
kindly allowing us to exceed our page allocation in describing the
WAVES concept.

\end{document}